\documentclass{aa}  
\usepackage{graphicx}
\usepackage{txfonts}
\usepackage{psfig}
\usepackage{epsfig}
\usepackage{natbib}
\bibpunct{(}{)}{;}{a}{}{,} 

\def \hcm {\hbox {\ifmmode $ cm$^{-2}\else cm$^{-2}$\fi}}

\def\approxgt{\mathrel{\hbox{\rlap{\lower.55ex \hbox {$\sim$}}
        \kern-.3em \raise.4ex \hbox{$>$}}}}
\def\approxlt{\mathrel{\hbox{\rlap{\lower.55ex \hbox {$\sim$}}
        \kern-.3em \raise.4ex \hbox{$<$}}}}

\begin{document}
   \title{The stellar content of low redshift radio galaxies from near-infrared spectroscopy}

   \subtitle{}

   \author{T. Hyv\"onen\inst{1}, J.K. Kotilainen\inst{1}, J. Reunanen\inst{1} \and R. Falomo\inst{2}}

   \offprints{T. Hyv\"onen}

   \institute{Tuorla Observatory, University of Turku, V\"ais\"al\"antie 20, FIN--21500 Piikki\"o, Finland\\
              \email{totahy@utu.fi, jarkot@utu.fi, reunanen@ftml.net}
	\and
             INAF -- Osservatorio Astronomico di Padova, Vicolo dell'Osservatorio 5, I-35122 Padova, Italy\\
             \email{falomo@pd.astro.it}
             }

   \date{Received; accepted}

 
  \abstract
   {We present medium spectral resolution near-infrared (NIR) $HK$-band spectra for eight low redshift ($z<0.06$) radio galaxies 
to study the NIR stellar properties of their host galaxies. The sample was selected from the radio galaxy sample imaged previously in the $B$- and $R$-band. 
They were found to be bluer than inactive elliptical galaxies, possibly indicating a recent star formation episode. As a homogeneous comparison sample, 
we used nine inactive elliptical galaxies that were observed with the same telescope and detector with similar resolution and 
wavelength range.}
   {The aim of the study is, by using the advantage of NIR absorption features, to compare the NIR spectral properties of radio galaxies to those of 
inactive early-type galaxies and, furthermore, produce the first NIR $HK$-band spectra for low redshift radio galaxies.}
   {For the radio galaxy and inactive elliptical samples, spectral indices of several diagnostic absorption features, namely 
SiI($1.589\mu$m), CO($1.619\mu$m) in the $H$-band and NaI($2.207\mu$m), CaI($2.263\mu$m), CO($>$$2.29\mu$m) 
in the $K$-band, were measured. The strength of absorption lines depends on the luminosity and/or temperature of stars and, 
therefore, spectral indices can be used to trace the stellar population of galaxies. To characterize the age of the populations, 
the measured EWs of the absorption features were fitted with the corresponding theoretical evolutionary curves of the EWs 
calculated by the stellar synthesis model.}
   {On average, EW(CO 2.29) of radio galaxies is somewhat greater than that of inactive ellipticals. Most likely, EW(CO 2.29) is 
not significantly affected by dilution, and thus indicating that elliptical galaxies containing AGN are in a different 
stage in their evolution than inactive ellipticals. This is also supported by comparing other NIR absorption line features, 
such as CaI and NaI, with each other. Based on the 
theoretical evolutionary curves of EWs absorption features are consistent with the intermediate age stellar population, suggesting that 
host galaxies contain both an old and intermediate age components. Intermediate age population is also 
consistent with previous optical spectroscopy studies, which have shown evidence on the intermediate age ($\sim$2 Gyr) stellar 
population of radio galaxies, and also in some of the early-type galaxies.}
   {Intermediate stellar population component indicates that radio galaxies have experienced a star formation epoch relatively recently. 
The existence of intermediate stellar population is a link between the star formation episode, possibly induced by interaction 
or merging event, and the triggering of the nuclear activity.}

   \keywords{galaxies: active -- galaxies: elliptical and lenticular, cD -- galaxies: general -- 
   galaxies: interactions -- galaxies: nuclei -- galaxies: stellar content}

   \titlerunning{The stellar population of low redshift radio galaxies}
   \authorrunning{Hyv\"onen et al.}

   \maketitle

\section{Introduction}

Recent observations have shown that most bulge-dominated galaxies contain a supermassive black 
hole in their center \citep[e.g.,][]{comb05}. This indicates that nuclear activity phenomenon 
can be a transient phase in evolution history of a galaxy that can be understood in terms 
of the central black hole fueled by in-falling gas toward the central region. The in-falling gas 
might be triggered either by merging processes of the spheroid formation at high redshift 
or during milder interaction of the galaxies at lower redshift \citep{cava00}. Dynamic interaction 
of an early-type galaxy with a gas rich late-type galaxy may often induce strong star 
formation as indicated by spectroscopic studies of local field early-type galaxies belonging 
to pairs or exhibiting shells \citep[e.g.,][]{long00,tant04} and theoretical studies of 
the merging process \citep[e.g.,][]{koji97}. Thus, the same mechanism may be responsible for 
both the nuclear activity and star formation. Characterizing the stellar populations in the central 
regions of galaxies, where supermassive black holes and recent star formation may co-exist, 
provides essential information for probing their origin and evolution. AGN feedback provides a mechanism to quench star 
formation on a short timescale \citep[e.g.,][]{spri05} and to replicate the observed 
bimodal color distribution of galaxies \citep[e.g.,][]{stra01}. The energy released by the AGN 
expels enough gas to quench both star formation and further black hole growth and, thus, 
determines the duration of starburst and AGN phase.

Most luminous AGN are hosted by large and luminous (massive) elliptical galaxies 
\citep[e.g.,][]{falo99,govo00,heid04,koti04,koti05,hyvo07a}. There is increasing 
evidence that the host galaxies of various types of AGN are bluer than inactive 
ellipticals \citep{govo00,scha00,orn03,sanc04,jahn04,koti04,hyvo07b} 
and they do not follow the color-magnitude relation of inactive ellipticals \citep{pele90}  
indicating a recent star formation episode. This scenario is also supported by optical spectroscopic studies of nearby 
radio galaxies (RG) \citep{raim05,holt07}. It is worth to note that 
the observed blue colors are not affected by the dust content of the galaxy, because in the 
presence of dust reddening the intrinsic color of the galaxy would be even bluer. 
Inactive elliptical galaxies have experienced a relatively short star formation episode at $z\sim$2 and evolved 
passively ever since \citep{carr07}. However, the distribution of the ages of stellar population of 
inactive ellipticals is inhomogeneous, often showing an intermediate age population ($\sim$3 Gyr), 
in addition to the energetically dominant, very old ($\sim$15 Gyr) populations \citep{yama06,silv08}. 
It is important to note that there is a downsizing effect, in the sense that the intermediate age populations are mostly found in early-type galaxies with low velocity dispersion, i.e. low mass, while the most massive galaxies are all very old. However,  there are also indications that galaxies in denser enviroments (clusters) are slightly older than field galaxies \citep{Trag00,kunt02}.
It is likely that the old 
stellar population is also dominating the stellar content of AGN hosts which has undergone a single recent 
star formation burst with only a small fraction of mass involved in the episode. The broad 
distribution of the colors most probably indicates an object-to-object difference in the age since the last 
star formation episode. The existence and the age of the young stellar 
population is a link between the interaction and/or merging event and the star formation episode 
and the triggering of the nuclear activity. Unfortunately, age estimates are not straightforward because of the metallicity effect. 
Strength of CO absorption lines of old and metal-rich stellar populations is similar to young starbursts with lower metallicity. 
Also, the cooler the younger -relation does not always hold because temperature depends on metallicity.

Luminosity of the majority of AGN is dominated by bright nuclear source contaminating the 
luminosity of the underlying host galaxy. Thus, spectroscopic observations of the 
stellar content of the host is difficult. Such studies 
\citep[e.g.,][]{boro85,oliv95,orig97,oliv99,moba00} have usually been limited by small heterogeneous samples, 
arrays and wavelength range. These studies have only recently been followed by systematic studies 
by e.g. \citet{nola01}, \citet{jahn02}, \citet{raim05} and \citet{silv08}. According to the unified model of radio-loud AGN, RGs are 
seen edge-on so that the bright central region is obscured making spectroscopy 
of the inner part of the host galaxy much easier. UV/optical spectroscopy of the inner 
region of RGs suggests that a large number of RGs contain a significant contribution 
from a young stellar population \citep[e.g.,][]{aret01,tadh02,will04,raim05,holt07}, probably 
related to star formation episode in the near past. Optical Lick spectral indices have 
usually been used in stellar population studies \citep{wort97}, but they have problems due to 
blended features and various population of galaxies. Furthermore, in the optical wavelengths, 
spectral features from the young stellar population are difficult to observe due to possible 
presence of other emission components, such as nebular continuum, and direct and scattered 
AGN emission. NIR spectroscopy has 
advantages compared to optical, because RGB stars dominate at $2\mu$m and 
the wavelength region contains many diagnostic stellar absorption 
lines (e.g., SiI $1.589\mu$m, CO(6-3) $1.619\mu$m, NaI $2.207\mu$m, CaI $2.263\mu$m 
and CO(2-0) bandhead $>$$2.29\mu$m) which can be used as indicators of stellar population in terms 
of their temperature and/or luminosity.

NIR spectroscopy has mostly been used to study high redshift RGs and there are only few studies 
for individual sources at low redshift \citep{bell03,tadh03}. 
We present NIR spectroscopy for eight nearby ($z<0.06$) RGs (one of them, PKS 0521-36 is also 
classified as BL Lac object) with reasonable spatial resolution ($\sim1$kpc) and medium spectral resolution 
($R\sim1000$) to study the NIR properties and ages of their stellar populations compared with 
the sample of nine inactive ellipticals. Except for individual sources \citep[e.g.,][]{tadh02} these 
are the first NIR spectra of a sizable sample of low redshift RGs. 
The RG sample was previously observed in $R$- and $B$-band by \citet{fasa96}, and 
detailed $B$--$R$ color information are available for them. 
The sample was selected among the objects of \citet{govo00} that are bluer than inactive elliptical galaxies 
and do not follow the $B$--$R$ color-magnitude relation of inactive ellipticals.
Long-slit spectra of RGs are 
extracted across the nucleus to detect all the important diagnostic absorption lines, and extended emission of 
the host galaxy can be studied to large distance from the nucleus. 
Our aim is to study the stellar population of RGs based on the NIR stellar absorption indices, 
and to compare the indices of RGs with inactive counterparts of similar morphology types. 
This will have important implications for understanding nuclear star formation histories and the evolution of RGs.

In Section 2, we describe the sample, observations, data reduction, and methods of analysis. 
In Section 3, we present the results and discussion concerning the properties of the galaxies. 
Summary and conclusions are given in Section 4.

\section{Observations, data reduction, and analysis}

The observations of the RGs were carried out in September 2005 at the 
ESO New Technology Telescope (NTT) using SOFI camera with pixel scale 0.288 arcsec pix$^{-1}$. The medium resolution ($R\sim1000$) 
long-slit $HK$-band spectra were taken across the nucleus with red grism and slit width 1.0 arcsec. The wavelength 
range of the spectra is from 1.5 to $2.5\mu$m with typical integration time of 1440 s.
The spectra of each target were taken with pairs using ABBA observing cycle and moving the target in the slit. 

Data reduction was performed using IRAF\footnote{IRAF is distributed 
by the National Optical Astronomy Observatories, which are operated by the Association of Universities for 
Research in Astronomy, Inc., under cooperative agreement with the National Science Foundation.}. 
Pairs were subtracted from each other to eliminate background sky emission and then divided by the 
flat-field image. Bad pixels and cosmic rays were masked out. Wavelength calibration was done using Xe-arc lamp 
calibration frames. Galaxy spectra were flux calibrated and divided by spectrophotometric standard star and then averaged. 
The nuclear (1.5 arcsec corresponding to $\sim$1.2 kpc in physical scale) and off-nuclear extended emission spectra were extracted for each target. In order to achieve sufficient S/N-ratio, the off-nuclear spectra were added together from both sides of 
the nucleus (1.5 arcsec/$\sim$1.2 kpc at each side). The $H$-band spectra were normalized by fitting the continuum 
level to both sides of the absorption features. In the 
$K$-band the continuum was fitted only to the shortward wavelengths from the CO bandhead. The journal of the RG sample 
is presented in Table~\ref{journal_rg}.

To have a reliable comparison between RGs and inactive ellipticals it is important that both samples have 
similar spectral resolution and aperture. As an inactive comparison sample 
we used the sample obtained and kindly provided for us by V. Ivanov (see also \citet{cese08}). The comparison sample consists of 
7 targets that were also observed by NTT/SOFI in March/April 2006 with an identical instrument configurations. These galaxies are 
classified as giant ellipticals with morphology type $T=-5$ but, 
in order to have morphology type $T\sim-3$ galaxies also in the comparison sample, we included two galaxies 
from the inactive early-type galaxy survey observed 
previously by us with NTT/SOFI using the same set of parameters and the reduction process as for the RG sample (Reunanen et al., in prep.). 
This data altogether gives us optimal spectral resolution matched sample for stellar population comparison. 
The journal of the inactive elliptical galaxy sample is presented in Table~\ref{journal_ell}. 

The equivalent widths (EW) of the spectral lines were derived by integrating over the lines of the normalized spectra. 
The integration limits for the particular line were the same for all targets.
The accuracy of the EWs is limited by the difficulty of determination of the continuum level 
in the vicinity of the particular line. Especially this holds for the $H$-band which is rich in spectral features, 
and for the CO(2-0) bandhead in the $K$-band because the continuum level has to be extrapolated longward of $2.295\mu$m. 
Furthermore, it is worth to keep in mind, that the stellar features are blended due to atomic and molecular features 
\citep{wall96}. 
There is no velocity dispersion data available for the RG sample, and we have 
assumed that they are roughly in agreement with those of inactive ellipticals 
used in this work. However, it is worth to note that the measured EWs of 
the indices are evidently upper limits due to spreading caused by 
the velocity dispersion. The possible dependence of the EWs on 
velocity dispersion and resolution was checked by degrading the medium 
resolution stellar spectra from \citet{wall97} to the lower resolution 
applicable for the RG sample, and measuring the indices of all the five 
diagnostic features for both resolutions. This comparison showed that 
the indices do not significantly depend on the degradation, in the sense that 
all the measured indices agree within the errors. 
In the normalized spectrum, a straight line was fit to the continuum points around the spectral feature. 
Central wavelength ranges for each feature were 1.585-1.593$\mu$m for SiI, 1.616-1.627$\mu$m for CO(6-3) and 
2.2924-$2.2977\mu$m for CO(2-0). For a direct comparison, the same wavelength ranges were adopted 
as those in \citet{orig93} and \citet{silv08}. The errors of the EWs are dominated by the uncertainties in fitting the continuum level and are 
typically $\sim$$0.5\AA$ in $H$-band and $K$-band metallic lines, and $\sim$$1.0\AA$ for the $K$-band CO(2-0) feature.

\setlength{\tabcolsep}{0.8mm}
\begin{table}
\centering
\caption{Journal of the radio galaxy sample.$^{\mathrm{a}}$}
\label{journal_rg}
\begin{tabular}{lllllll}
\hline
\noalign{\smallskip}
Target & $z$ & T & $B-R$ & $M_K$ & $T_{exp}$ & Date\\
& & & & mag & min & \\
(1) & (2) & (3) & (4) & (5) & (6) & (7)\\
\noalign{\smallskip}
\hline
\noalign{\smallskip}
PKS 0023-33  & 0.0498 & -5   & 1.47 & 11.3 & 32 &27/09/2005\\
3C 29        & 0.0450 & -     & 1.57 & 11.0 & 24 &26/09/2005\\
NGC 612      & 0.0298 & -1.2 & 1.59 & 10.1 & 24 &25/09/2005\\
ESO 552-G14  & 0.0317 & -2.7 & 1.56 & 10.6 & 24 &27/09/2005\\
PKS 0521-36  & 0.0553 & -    &1.48  &11.0 &32 &25/09/2005\\
ESO 528-G36  & 0.0408 & -5 & 1.48 & 10.8 & 24 &27/09/2005\\
NGC 6998     & 0.0397 & -5 & 1.84 & 10.8 & 32&26/09/2005\\
PKS 2158-380 & 0.0333 & - &1.13 & 11.4 & 32 &27/09/2005\\
\noalign{\smallskip}
\hline
\end{tabular}
\begin{list}{}{}
\item[$^{\mathrm{a}}$] 
Column (1) gives the name of the target; (2) redshift; (3) morphology type; (4) the $B-R$ colour \citep{fasa96}; 
(5) $K$-band magnitude from 2MASS 
in 5 arcsec radius aperture; (6) integration time, and (7) the date of the observation.
\end{list}
\end{table}
\setlength{\tabcolsep}{1.5mm}

\setlength{\tabcolsep}{0.8mm}
\begin{table}
\centering
\caption{Journal of the elliptical galaxy sample.$^{\mathrm{a}}$}
\label{journal_ell}
\begin{tabular}{lllll}
\hline
\noalign{\smallskip}
Target & $z$ & T & $T_{exp}$ & Date\\
& & & s &\\
(1) & (2) & (3) & (4) & (5)\\
\noalign{\smallskip}
\hline
\noalign{\smallskip}
NGC 2640 & 0.00350 & -3 & 32 & 09/03/2006\\
NGC 4478 & 0.00469 & -5 & 40 & 17/04/2006\\
NGC 4472 & 0.00333 & -5 & 33 & 17/04/2006\\
NGC 4546 & 0.00350 & -3 & 32 & 11/03/2006\\ 
NGC 4564 & 0.00377 & -5 & 30 & 17/04/2006\\
NGC 4621 & 0.00143 & -5 & 20 & 18/04/2006\\
NGC 4649 & 0.00365 & -5 & 33 & 17/04/2006\\
NGC 4697 & 0.00415 & -5 & 23 & 17/04/2006\\
NGC 5576 & 0.00505 & -5 & 25 & 17/04/2006\\
\noalign{\smallskip}
\hline
\end{tabular}
\begin{list}{}{}
\item[$^{\mathrm{a}}$] 
Column (1) gives the name of the target; (2) redshift; (3) morphology type; 
(4) integration time, and (5) the date of the observation.
\end{list}
\end{table}
\setlength{\tabcolsep}{1.5mm}

\section{Results}

\begin{figure*}
\centering
\includegraphics[width=17cm,height=15cm]{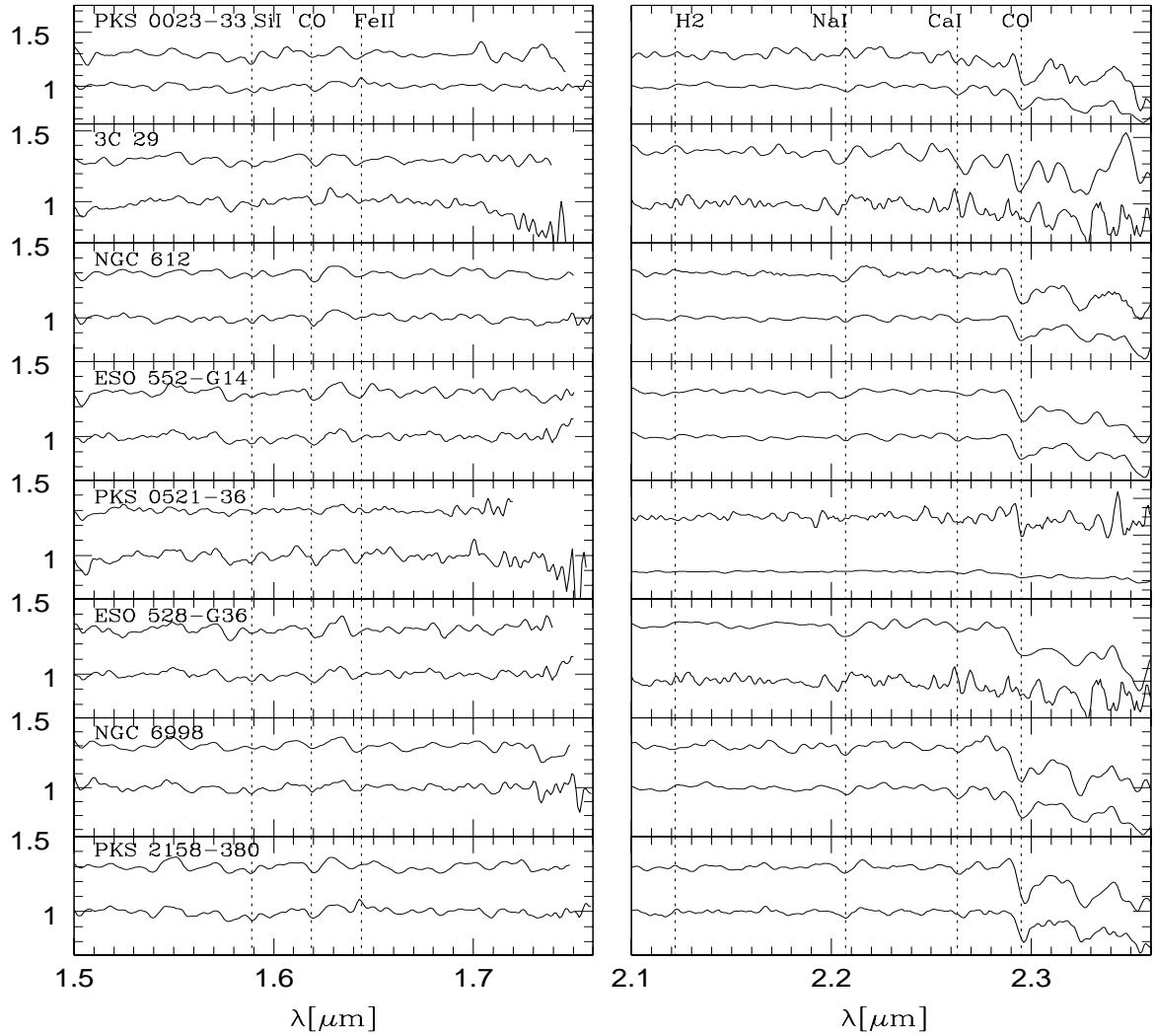}
\caption[]{The $HK$-band off-nuclear (upper) and nuclear (lower) spectra of radio galaxies. 
The most prominent diagnostic spectral absorption features together with supernova activity indicator emission lines (H$_2$ and [FeII]) 
are marked by dotted vertical lines.
\label{spectra}}
\end{figure*}

The $HK$-band off-nuclear and nuclear spectra for RG sample are presented in Fig.~\ref{spectra}. 
In all off-nuclear and nuclear spectra of RGs, $H$-band spectral absorption features 
from stellar population (SiI, CO(6-3)) are clearly visible as are in the most cases the features NaI, CaI, CO(2-0) 
in the $K$-band. The EWs of all diagnostic spectral features can be measured for all inactive elliptical galaxies, except SiI in two cases 
(NGC  4472, NGC 4564). The EWs of the absorption lines for RGs and inactive ellipticals are presented in Table~\ref{ew}. 

\begin{figure*}
\centering
\includegraphics[width=17cm]{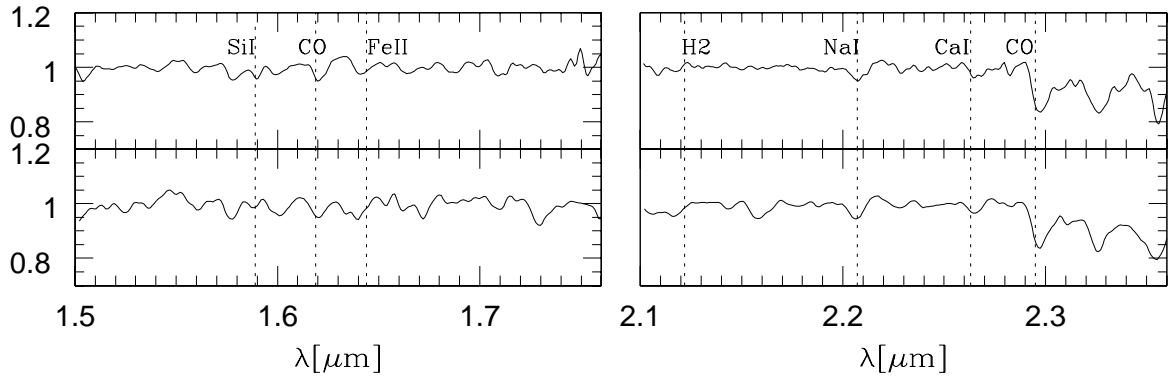}
\caption[]{The $HK$-band composite spectra for radio galaxy (upper panels) and inactive elliptical galaxy (lower panels) 
samples. The most prominent diagnostic spectral features are marked by dotted vertical lines.
\label{comb_spectra}}
\end{figure*}

By combining each individual redshift corrected spectra together, we also created a representative average spectrum for 
RGs and inactive ellipticals (PKS 0521-36 was not included in the composite spectrum of RGs) which are shown 
in Fig.~\ref{comb_spectra}. These spectra can find applications as template spectra for the underlying galaxy 
which can be subtracted from the spectra of composite stellar systems (i.e. galaxies hosting an AGN).

\begin{table}
\caption{EWs of the absorption lines for off-nuclear and nuclear spectra of radio galaxies and inactive 
elliptical galaxies in units of $\AA$.}
\label{ew}
\begin{tabular}{lllllll}
\hline
\noalign{\smallskip}
Target & T & SiI & CO(6-3) & NaI & CaI & CO(2-0) \\
$\lambda$ [$\mu$m]& & 1.589 & 1.619 & 2.207 & 2.263 & $>2.295$ \\
\noalign{\smallskip}
\hline
\noalign{\smallskip}
RGs: off-nuclear & & & & & & \\
PKS 0023-33  & -5   & 5.3 & 5.5 & -    & 3.7 & 17.1 \\
3C 29        & -     & 4.5 & 4.6 & 4.2 & 3.5 & 20.5 \\
NGC 612      & -1.2 & 2.6 & 4.5 & 3.4 & 3.0 & 18.8 \\
ESO 552-G14  & -2.7 & 3.4 & 4.5 & 4.4 & 2.9 & 15.1 \\
PKS 0521-36  &  -    & 1.2 & 2.5 & 1.1 & 2.5 &  5.4 \\
ESO 528-G36  & -5   & 2.3 & 3.3 & 4.4 & 3.1 & 20.7 \\
NGC 6998     & -5   & 2.8 & 4.2 & 3.9 & 3.8 & 29.0 \\
PKS 2158-380 &  -    & 4.8 & 3.5 & 4.9 & 3.3 & 20.6 \\
             &      &     &     &     &     &      \\
RGs: nuclear   &     &           &      \\
PKS 0023-33  & -5   & 3.1 & 6.0 & 4.5 & 2.9 & 17.5 \\
3C 29        &  -    & 2.5 & 3.8 &  -   & -    & -     \\
NGC 612      & -1.2 & 2.6 & 3.8 & 3.9 & 3.4 & 20.1 \\
ESO 552-G14  & -2.7 & 2.3 & 5.1 & 2.5 & 2.3 & 14.9 \\
PKS 0521-36  & & 0.8 & 3.1 &- & - & -  \\  
ESO 528-G36  & -5   & 1.8 & 5.8 & 3.7 & 7.7 & 13.3 \\
NGC 6998     & -5   & 2.3 & 3.3 & 5.3 & 5.8 & 16.5 \\
PKS 2158-380 &  -    & 3.8 & 5.1 & 3.9 & 3.4 & 14.6 \\
             &      &     &     &     &     &      \\
Ellipticals: &      &     &     &     &     &      \\
NGC 2640     & -3   & 5.1 & 3.1 & 5.1 & 3.1 & 20.7 \\
NGC 4478     & -5   & 1.7 & 5.0 & 3.5 & 3.7 & 12.5 \\
NGC 4472     & -5   &  -   & 4.5 & 5.5 & 2.7 & 17.4 \\
NGC 4546     & -3   & 4.8 & 4.5 & 4.5 & 2.5 & 17.3 \\
NGC 4564     & -5   &  -   & 4.0 & 5.5 & 2.8 & 17.4 \\
NGC 4621     & -5   & 2.5 & 6.5 & 5.2 & 5.6 & 16.6 \\
NGC 4649     & -5   & 2.1 & 5.1 & 4.9 & 2.7 & 17.2 \\
NGC 4697     & -5   & 2.1 & 5.5 & 4.1 & 3.5 & 16.3 \\
NGC 5576     & -5   & 2.2 & 5.9 & 4.0 & 2.4 & 14.2 \\
\noalign{\smallskip}
\hline
\end{tabular}
\end{table}
\setlength{\tabcolsep}{1.5mm}

\subsection{Near-infrared line indices as a stellar population indicators}

NIR spectra can be used as a tool for studying stellar population in galaxies and star clusters, 
because the spectral domain contains many strong stellar absorption features. Especially, 
the $K$-band has been widely studied \citep[e.g.,][]{merr79,klei86,wall97}, and it is a natural choice 
for studying reddened (dusty) galaxies at low redshift \citep[e.g.,][]{moba00,manu01}. However, $H$-band  is even more suitable 
for population studies because in this band non-stellar emission, mostly caused by dust, has smaller effect 
than in the $K$-band but $H$-band domain has been studied systematically only relatively recently
\citep[e.g.,][]{orig93,oliv95,frem07}.

The most evident absorption features in the $K$-band are the molecular CO(2-0) 
bandhead ($>$$2.295\mu$m), atomic Br-$\gamma$ ($2.166\mu$m), NaI ($2.207\mu$m), 
and CaI ($2.263\mu$m) lines. Br-$\gamma$ is the strongest absorption feature within stars earlier 
than K5, but becomes undetectable in stars later than K5 \citep{klei86}, while 
NaI is the strongest atomic feature when effective temperature is $T_{eff}<3400 K$ \citep{ali95}. 
$H$-band has a very complex absorption line pattern due to a number of metallic and molecular lines, 
especially in cool stars. The most studied $H$-band features are atomic SiI ($1.589\mu$m) and 
molecular CO(6-3) ($1.619\mu$m) \citep[e.g.,][]{orig93}. 
It is well known, that the NaI, CaI and CO features in K and M giant stars become stronger with redder $J$--$K$ color 
\citep{rami97,fors00,frog01}. At a given $J$--$K$, giants in more metal rich clusters have stronger 
NaI and CaI features \citep{frog01}. In the NIR, the integrated light is dominated by stars that are 
very different from those that contribute significantly at visible wavelengths. The first overtone 
$^{12}$CO bandheads are the most prominent features in the NIR spectra of old and intermediate 
age populations, and so are potentially prime diagnostics of stellar content. The depth of these 
features are predicted to be sensitive to population parameters \citep[e.g.,][]{vazq03}. 
NaI and CaI indices sample transitions from more than one element, complicating the sensitivity 
of these lines to chemical abundance. Such contamination is also common among optical line indices \citep[e.g.,][]{wort94}. 
Whereas the integrated light at visible wavelengths comes from a mix of stellar types, the $K$-band 
light from all but the youngest stellar systems is dominated by evolved giants and AGB (Asymptotic Giant Branch) stars, whose 
photometric and spectroscopic properties are sensitive to age and metallicity \citep{frog78,mara05}. 
The relative strengths of the NIR features can be used to compare the stellar contents of the 
central regions of galaxies, and to provide insights into their history.

Currently there are no self-consistent theoretical spectral synthesis models for the interpretation 
of NIR spectra of galaxies. Thus, one has to rely on e.g. their comparison with high-resolution 
NIR stellar spectral atlases \citep[e.g.,][]{wall96}. 
The usefulness of stellar absorption lines as tools for tracing the stellar population of galaxies is based on the 
fact that particular lines are sensitive to 
the spectral type (temperature) and/or luminosity (gravity) of the star \citep{klei86,ali95,rami97}. 
CO lines decrease with increasing temperature 
due to variation of the CO/C abundance ratio, being detectable at temperatures lower than $T_{eff}\sim3500$K and 
increasing with luminosity. Thus, $K$-band CO bandhead is very strong in young giant and supergiant 
stars (10 Myr-100 Myr) and strong in cool asymptotic giant branch (AGB) stars (100 Myr-1 Gyr; \citet{oliv95}), 
while it is weaker in older population. This makes CO bandhead suitable to trace recent star formation 
in galaxies \citep{mayy97}. 
However, some line ratios, such as EW(CO 1.62)/EW(SiI 1.59) and EW(CO 1.62)/EW(CO 2.29), 
are even better temperature indicators than single lines \citep{orig93}. Especially, the former ratio 
is useful because the two lines have very similar wavelengths and therefore the ratio is not sensitive to dilution of the 
emission of the hot gas.

\subsection{Near-infrared line indices of radio galaxies}

\begin{figure}
\centering
\includegraphics[width=9cm]{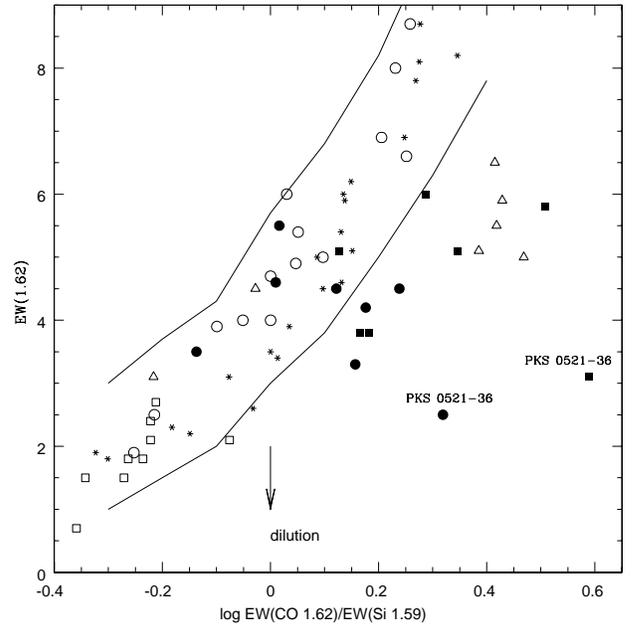}
\caption[]{EW(CO 1.62) as a function of log[EW(CO 1.62)/(EW(Si 1.59)] for 
off-nuclear spectra of radio galaxies (filled circles), nuclear spectra of radio galaxies (filled squares) and 
spectra of inactive elliptical galaxies (open triangles). Different 
luminosity types of stars are shown as symbols: supergiants I (open circles), 
giants III (asterisks) from \citet{rami97} and dwarfs V (open squares) from \citet{ali95}. 
The lines enclose the area occupied by stars with no dilution and 
the arrow gives the direction of the effects of dilution.
\label{co162si159}}
\end{figure}

\begin{figure}
\centering
\includegraphics[width=9cm]{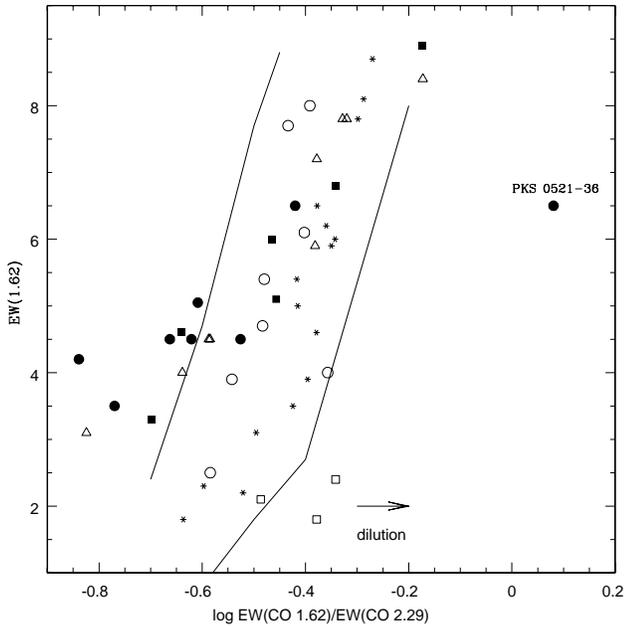}
\caption[]{EW(CO 1.62) as a function of log[EW(CO 1.62)/(EW(CO 2.29)]. The meaning of the symbols is the same as in 
Fig.~\ref{co162si159}.
\label{co162co229}}
\end{figure}

Non-stellar thermal, such as dust surrounding young star forming regions, and/or non-thermal dilution 
reduce the intrinsic values of EWs originating from the stellar population. 
However, the dust observed in NIR is close to its sublimation temperature, and in such cases HII 
emission line series are expected to be observed. The comparison of absorption indices in different filters 
depends on the level of dilution by the non-stellar component that can be different in $H$- and $K$-band. 
The dilution fraction of the continuum emission at $1.6\mu$m can be 
estimated from the plot of EW(CO 1.62) vs. EW(CO 1.62)/EW(Si 1.59) which is shown in Fig.~\ref{co162si159}. 
Objects with diluted stellar features lie below the locus occupied by stars, because 
they have a shallower CO(1.62) index, while EW(CO 1.62)/EW(Si 1.59) is not significantly affected 
by dilution since two the features lie close in wavelength. The fraction of non-stellar continuum is simply 
given by vertical displacement of the point in the diagram. 
The dilution at $2.3\mu$m can be determined in a similar way from the plot of EW(CO 1.62) vs. EW(CO 1.62)/EW(CO 2.29) 
shown in Fig.~\ref{co162co229}. 
In that case, objects with significant non-stellar continua 
lie on the right hand side from the enclosed region. Note, that one first needs to correct EW(CO 1.62) for dilution before determining 
the non-stellar fraction from the horizontal displacement in the diagram. 
In the $H$- and $K$-band, all off-nuclear RGs have either no or only slight dilution, 
while the BL Lac object (PKS 0521-36) has significant dilution in both bands reducing the measured EWs. 
Its large non-thermal dilution is expected in terms of powerful jet emission 
from the BL Lac object. None of the inactive elliptical galaxies show any dilution in the $K$-band indicating that 
the EWs of RGs and ellipticals are not significantly affected by non-stellar dilution. Another estimate of 
dilution can be obtained from comparison of nuclear and off-nuclear continuum shapes, in which case redder 
spectrum is expected in nucleus. The comparison was done between off-nuclear and nuclear spectra of RGs, but there was no 
difference between the spectra. This indicates that no significant amount of dust is 
present in the nuclear region of RGs and, as a consequence, HII emission lines are not seen in the RG spectra.

Table~\ref{ave_ew} gives the average values of off-nuclear and nuclear EWs of SiI, CO(1.62), NaI, CaI and CO(2.29) 
absorption lines for RGs and the corresponding values for inactive ellipticals. 
While the off-nuclear EWs of RGs are generally in agreement with the nuclear EWs of RGs and those of ellipticals, 
the off-nuclear EW(CO 2.29) of RGs is larger than that in the nuclear region of RGs and in elliptical galaxies.
As was previously mentioned, it is unlikely that this 
difference is caused by non-stellar dilution. The EW(CO 2.29) is sensitive to the luminosity class of stars 
being smaller in dwarfs than in giant stars. The difference of EW(CO 2.29) 
between RGs and inactive ellipticals may indicate a different stellar populations.

\begin{figure}
\includegraphics[width=9cm]{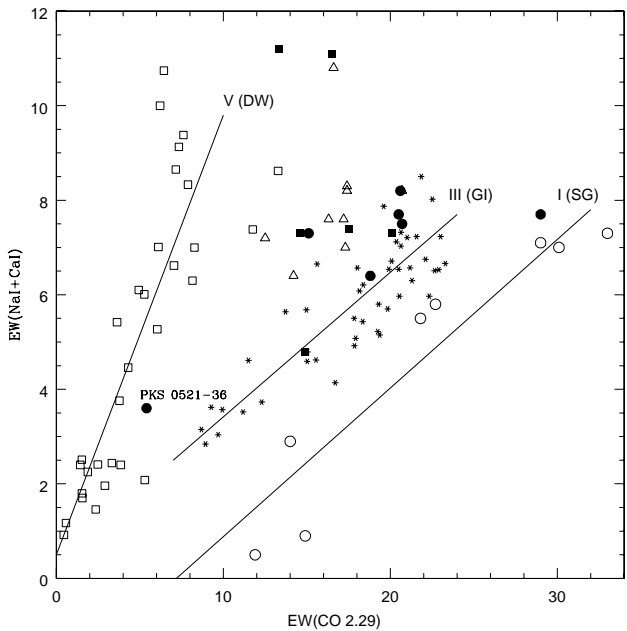}
\caption[]{The sum of the EWs of NaI and CaI as a function of EW(CO 2.29). 
The lines show the loci of different luminosity types of stars. The meaning of the symbols is the same as in Fig.~\ref{co162si159}.
\label{naca}}
\end{figure}

\begin{figure}
\includegraphics[width=9cm]{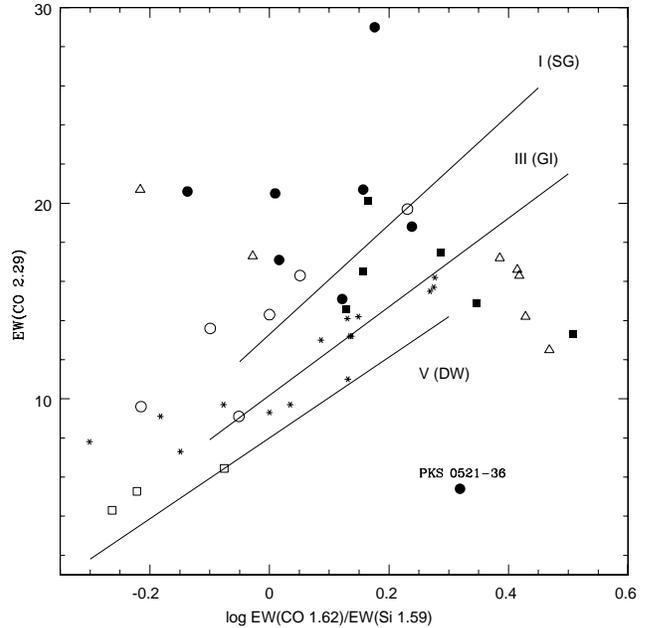}
\caption[]{EW(CO 2.29) as a function of log[EW(CO 1.62)/(EW(Si 1.59)]. The meanings of the symbols and the lines are the same as in 
Fig.~\ref{co162si159} and Fig.~\ref{naca}.
\label{co162si159co229}}
\end{figure}

Figure~\ref{naca} shows the EW(NaI+CaI) as a function of EW(CO 2.29) for off-nuclear and nuclear RGs, 
and inactive elliptical 
galaxies together with stars of different spectral type. 
Caution should be exercised when comparing stellar and galactic data, as galaxies are composite 
stellar systems and their spectra show the integrated contributions from stars spanning a range of 
properties. Still, given that the NIR spectral region is dominated by evolved (RGB) stars, this issue is not 
as critical as at visible wavelengths, where stars contribute over a much larger range of evolutionary states 
(both RGB and main sequence stars e.g. \citet{mara05}). All three luminosity 
classes of stars have a relatively tight linear correlation between EW(NaI+CaI) and EW(CO 2.29). 
Most of the off-nuclear RGs are in good agreement with the relation of 
giant (III) stars, while inactive ellipticals deviate more strongly from that relation. 
On the other hand, the plot of EW(CO 2.29) vs. EW(CO 1.62)/EW(Si 1.59) (Fig.~\ref{co162si159co229}) indicates 
that the stellar populations are mostly dominated by supergiants (I). However, it is possible that the CO(1.62) line is 
underestimated or the SiI line overestimated decreasing the line ratio of EW(SiI)/EW(CO 1.62) and thus, in both cases, 
shifting the real data points to the left in Fig.~\ref{co162si159co229}. It is 
reasonable that SiI line is slightly overestimated, because the absorption feature is located in the $H$-band which has 
numerous absorption lines and, in medium resolution spectra, the lines are blended. 
In conclusion, the majority of RGs are consistent with the relation of giant stars. Only a few RGs are readily 
in agreement with ellipticals, while the majority are displaced toward giants and supergiants 
indicating the difference in their stellar populations. 

Almost all RGs and inactive ellipticals have EW(CO 2.29)$>15\AA$ which is larger than the EW of 
K giants (from $10\AA$  to $15\AA$) or main sequence stars. In fact, the only stars with EW of CO(2.29) higher 
than $15\AA$ are M giants and K and M supergiants. The stellar population of these galaxies must 
therefore be dominated by very cool stars, probably M giants, to match the measured EWs. Note, however, that since 
M stars are much brighter in the $K$-band than K stars, this does not imply that M giants are the only significant 
luminosity-weighted population.

Star formation in the vicinity of nucleus can be activated by nuclear activity. Indeed, there 
is evidence for ongoing circumnuclear star formation in the central region of the BL Lac PKS 2005-489 
\citep{bres06}. Different star formation rates within a galaxy might set up a radial stellar population gradient 
which can be seen as a difference in line strengths between nuclear and off-nuclear 
spectra. However, non-stellar dilution must be taken into account that weakens the absorption lines of stellar 
population in the nuclear region. Only 3C 29 shows a systematic variation in the 
strength of the lines between off-nuclear and nuclear spectra. The variation of the line strengths of 3C 29 can be due to a 
stellar population gradient but, however, deeper and better spatial resolution 
spectra are needed to confirm the possible gradient.

\begin{figure}
\includegraphics[width=16cm]{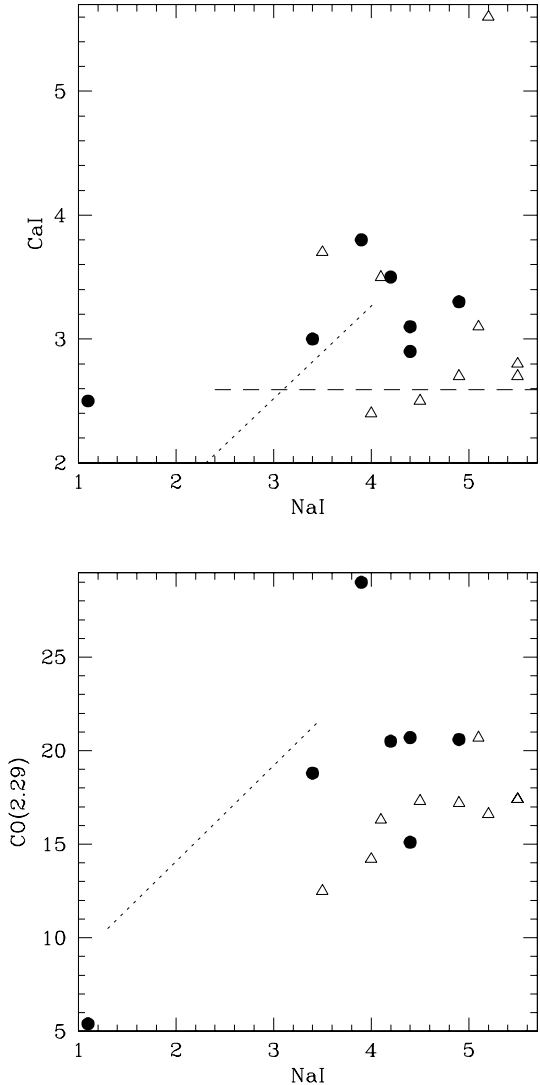}
\caption[]{EWs of CaI and CO(2.29) against EW(NaI). Symbols are the same as in 
Fig.~\ref{co162si159}. The dotted and long-dashed lines represent the relations of the solar metallicity 
cluster star fits and purely old population \citep{silv08}, respectively.
\label{ew_rg_new}}
\end{figure}

EW(CaI) and EW(CO 2.29) of off-nuclear RGs and elliptical galaxies are plotted against EW(NaI) in Fig.~\ref{ew_rg_new}. 
They are compared with the relation of solar metallicity cluster stars and purely old population galaxies adapted from \citet{silv08}. 
In the plot of CaI vs. NaI, majority of the ellipticals occupy the locus which is consistent with 
the relation of purely old galaxies, whereas all RGs have larger CaI indices. Furthermore, in both 
EW(CaI) and EW(CO 2.29) plots, majority of the RGs and ellipticals also have larger NaI 
indices than those of solar metallicity cluster stars, being consistent with the results obtained by \citet{silv08} 
for early-type galaxies of Fornax cluster. 
Unlike found by \citet{cese08}, our elliptical galaxy sample do not show a significant correlation between EW(CaI) and EW(NaI). 
In both plots, RGs deviates from the ellipticals suggesting that elliptical galaxies with AGN are in a different stage in 
their evolution than inactive counterparts.


\setlength{\tabcolsep}{0.8mm}
\begin{table}
\caption{Average values of EWs for off-nuclear and nuclear radio galaxies and inactive elliptical galaxies compared to 
EWs of supergiants (SG), giants (G) and dwarfs (DW).}
\label{ave_ew}
\begin{tabular}{llllll}
\hline
\noalign{\smallskip}
&SiI & CO(6-3) & NaI & CaI & CO(2-0) \\
\noalign{\smallskip}
\hline
\noalign{\smallskip}
RG: off-nuc & $3.7\pm1.2$ & $4.3\pm0.7$ & $4.2\pm0.5$ & $3.3\pm0.3$ & $20.3\pm4.4$ \\
RG: nuc & $2.6\pm0.7$ & $4.7\pm1.1$ & $4.0\pm0.9$ & $4.3\pm2.1$ & $16.2\pm2.4$ \\
Ellipticals: & $2.9\pm1.4$ & $4.9\pm1.0$ & $4.7\pm0.7$ & $3.2\pm1.0$ & $16.6\pm2.3$ \\
Supergiants (I):      & 4.3$\pm$0.6 & 4.6$\pm$2.7 &- &- & 17.6$\pm$8.0 \\
Giants  (III):    & 3.8$\pm$0.4 & $5.1\pm1.9$ & $2.8\pm0.5$ & $2.9\pm0.8$ & $16.4\pm4.7$ \\
Dwarfs (V):      & 3.2$\pm$0.8 & $1.8\pm0.6$ & $2.9\pm1.9$ & $2.6\pm1.4$ & $5.3\pm3.1$ \\
\noalign{\smallskip}
\hline
\end{tabular}
\end{table}
\setlength{\tabcolsep}{1.5mm}

\subsection{Age estimates for stellar population of radio galaxies}

It is well known that the majority of inactive elliptical galaxies have not 
experienced recent significant star formation, but 
are dominated by passively evolved relatively old population of late-type giant RGB stars. However, many recent imaging 
\citep[e.g.,][]{govo00,hyvo07b} and spectroscopic studies \citep[e.g.,][]{raim05} have shown that, unlike 
inactive ellipticals, RGs have experienced a recent star formation episode and, thus, bluer 
host galaxy colors are due to younger stellar population component. \citet{raim05} obtained spectra for 24 RGs at optical 
wavelengths, and they found a systematic difference in stellar population between RGs 
and inactive ellipticals, in the sense that RGs have a larger contribution from an intermediate age population.

Although stellar population synthesis models \citep[e.g.,][]{fioc97,bruz03} at NIR wavelengths are 
not as well developed as at visible wavelengths, they can be used to provide 
estimates for the ages of the stellar population of galaxies \citep{riff08}. Theoretical spectral 
fitting is not currently possible at NIR, but synthesis models can be 
used to derive the time evolution of the EWs of diagnostic stellar lines parametrized by IMF 
and the metallicity of stellar population.

To calculate the evolution of EW for the NIR absorption lines 
we used the model of \citet{mara05} with Salpeter IMF, solar metallicity and 
SSP model. That model was preferred over the continuous star formation model 
because the latter predicts Br-$\gamma$ lines which are not seen in our RG spectra and, furthermore, in previous 
studies \citep{kenn98,bend04} the observed spectra were found to be more consistent with the SSP model. Metallicity and age of stellar populations are degenerate, in the sense that increasing metallicity 
generates older stellar populations. However, very low metallicities of RGs and ellipticals are very 
unlikely because the observed line indices are significantly greater than those predicted by the low metallity models.

The measured EWs of each absorption line of off-nuclear RGs and inactive elliptical galaxies were compared with corresponding 
EW vs. time theoretical curve computed by the synthesis model, allowing to estimate the ages 
of the stellar population. 
Note that before this comparison, the observed spectra ($\sim$30 \AA) were smoothed to the same 
resolution of the Maraston05 model (100 \AA).
Theoretical curves were calculated using exactly the same parameters as those in a recent 
NIR spectroscopic study of starburst galaxies by \citet{riff08}, and we refer to their Fig. 4 for the behavior 
of the absorption lines as a function of age. 
Fitting was performed for $K$-band CaI feature and CO(2.29) bandhead, and the results of the fitting are given in Table~\ref{ages}. 
Note that the theoretical EW values are not unique with evolutionary time, but the same EW can have two to four 
different time values between very young ($\sim$10 Myr) and intermediate age ($\sim$2 Gyr) stellar population fits. 
EWs of CaI and CO(2.29) absorption lines of RGs are formally consistent with the fits of a young ($\sim$400 Myr) and 
intermediate age ($\sim$2 Gyr) stellar population. However, the young population is unlikely because massive stars 
passing the red giant phase will explode as supernovae, and emission lines of [FeII]($1.644\mu$m) and H$_2$1-0S(1)($2.212\mu$m) 
from the supernova 
should be seen. The lack of these lines in the observed spectra implies either weak or nonexistent supernova activity 
and, thus, nonexistent young population. For these reasons, although both are formally consistent with the observations, the intermediate age population is preferred 
over the young population in Table~\ref{ages}.

This result is in good agreement with the ages of nearby RGs derived from the optical spectra \citep{raim05}, and 
suggests the widespread existence of a two-component stellar population containing both an old dominating component and 
an intermediate age component, in both RGs and inactive ellipticals. Note that a similar two-component model was 
recently suggested by \citet{silv08} for three early-type galaxies of Fornax cluster. The sample of \citet{raim05} 
includes one common object to us, namely NGC 612. For that object, they estimated 
the stellar population to be $\sim2.4$ Gyr of age that is consistent with our result (Table~\ref{ages}). 
These ages are, on the other hand, only in reasonable agreement with those based on fitting single-age, single-metallicity stellar populations 
to optical and NIR colours. For example, using this method, \citet{jame06} derived luminosity-weighted 
mean ages between 3 and 14 Gyr, with a mean of $\sim$8 Gyr, for field and Virgo cluster elliptical galaxies. 
However, we note that near-infrared is much more sensitive spectral region to trace evolved RGB stars than optical, in the sense that even a relatively small RGB fraction in a stellar population generates strong contribution to the strength of near-infrared absorption features. This causes a systematic effect between the age estimates from optical and near-infrared methods. 
Although RGs and inactive ellipticals occupy different loci in the index vs. index plots of 
Fig.~\ref{ew_rg_new}, reflecting differencies in their stellar content, the best-fit age estimates of both 
samples are in reasonable agreement. Most likely, this indicates different metallicities in RGs and 
inactive ellipticals.

RGs appear to be slightly older than late-type starburst galaxies ($\sim$1 Gyr) observed 
by \citet{riff08} in NIR. This might indicate that RGs have experienced nuclear activity and enhanced star formation earlier than 
starburst galaxies, and in those RGs star formation have finished possibly due to AGN feedback phenomenon.

\begin{table}
\centering
\caption{Age estimates in Myr for the stellar population component of each off-nuclear radio galaxy and inactive elliptical galaxy together 
with the stellar population age of the combined RG and inactive elliptical samples.$^{\mathrm{a}}$}
\label{ages}
\begin{tabular}{l|ll|ll|ll}
\hline
\noalign{\smallskip}
Target & CaI & & CO &(2.29) & Young&Interm.\\
\noalign{\smallskip}
\hline
\noalign{\smallskip}
Radio Galaxies:& & & & & & \\
PKS 0023-33  & 350 & 1580 & 280  & 2000 & 315 & 1800 \\
3C 29        & 340 & 1800 & 680  & 1260 & 510 & 1500 \\
NGC 612      & 320 & 2000 & 630  & 1580 & 480 & 1800 \\
ESO 552-G14  & 320 & 2000 & 320  & 2630 & 320 & 2300 \\
PKS 0521-36  & 280 & 2240 &  -   & -    & 280 & 2200 \\
ESO 528-G36  & 310 & 2000 & 700  & 1300 & 510 & 1650\\
NGC 6998     & 350 & 1600 & -    & -    & 350 & 1600\\
PKS 2158-380 & 330 & 1800 & 700  & 1300 & 500 & 1500\\
Combined RG  & 350 & 2000 & 660  & 1400 & 500 & 1700\\
             &     &      &      &      &      \\
Ellipticals: &     &      &      &      &      \\
NGC 2640     & 320 & 2000 & 700  & 1300 & 510 & 1650 \\
NGC 4478     & 360 & 1600 & 270  & -    & 320 & 1600\\
NGC 4472     & 280 & 2140 & 400  & 2000 & 340 & 2100\\
NGC 4546     & 260 & 2240 & 380  & 2040 & 320 & 2100\\
NGC 4564     & 280 & 2140 & 400  & 2000 & 340 & 2100\\
NGC 4621     & 630 & 1100 & 370  & 2400 & 500 & 1750\\
NGC 4649     & 280 & 2100 & 370  & 2100 & 330 & 2100\\
NGC 4697     & 340 & 1800 & 360  & 2500 & 350 & 2150\\
NGC 5576     & 260 & 2300 & 320  & 2800 & 290 & 2550\\
Combined E   & 340 & 1900 & 350  & 2200 & 350 & 2050 \\
\noalign{\smallskip}
\hline
\end{tabular}
\begin{list}{}{}
\item[$^{\mathrm{a}}$]
Column (1) gives the name of target; (2) and (3) best-fit young and intermediate ages based on the EWs of CaI; (4) and (5) ages based on the 
EWs of CO(2.29), and (6) and (7) the average age of the young and intermediate age stellar population component. 
\end{list}
\end{table}
\setlength{\tabcolsep}{1.5mm}

\section{Conclusions}

We have observed the first NIR $HK$-band spectra of a sizable sample of low redshift RGs 
to characterize the EWs of their diagnostic NIR absorption line features and, in addition, based on the 
evolution models of the diagnostic features trace the intermediate age stellar population component. RG sample 
was compared with nearly similar size inactive elliptical galaxy sample obtained with the same telescope, 
instrument and wavelength range. For both samples spectral indices of several absorption features were measured in both bands, 
namely SiI ($1.589\mu$m), CO(6-3) ($1.619\mu$m) in the $H$-band, and NaI ($2.207\mu$m), CaI ($2.263\mu$m), 
CO(2-0) ($>$$2.29\mu$m) in the $K$-band. These lines are sensitive to the stellar temperature and/or 
luminosity so they can be used as a tool for identifying different spectral and luminosity classes of stellar population. 
NIR spectroscopy is powerful to study intermediate age stellar populations having advantages over 
optical spectroscopy to trace TP-AGB stars. 
Although currently stellar 
population synthesis models are not fully self-consistent for the interpretation of NIR spectra of 
galaxies, they still can be used to estimate 
the age of stellar population by comparing the measured EWs of diagnostic stellar lines to the EWs calculated by 
theoretical evolution curve. 

Plotting EW(NaI+CaI) vs.EW(CO 2.29) RGs are consistent with the relation of giants stars, whereas inactive ellipticals deviates more strongly from 
the relation toward main sequence dwarf stars, and on average, EW(CO2.29) of RGs is somewhat greater than that of inactive ellipticals. 
Most likely, EWs are not significantly affected by non-stellar dilution, 
and thus it indicates that elliptical galaxies containing AGN are in a different stage in their evolution than their inactive 
elliptical counterparts. Our result supports previous optical 
spectroscopic studies of RGs which have shown evidence from intermediate age stellar population component, 
indicating a link between a nuclear activity and star formation in the host galaxy. Furthermore, RGs appear to 
contain a slightly older stellar population component than those of late-type starburst galaxies suggesting 
that RGs have experienced enhanced star formation earlier than starburst galaxies. There are no H$_2$1-0S(1) and [FeII] 
emission lines in the RG spectra indicating either only very small or non-existing contribution from very young stellar population.

To have a better understanding of the stellar content of RGs, larger NIR spectroscopic sample 
is needed with higher spectral and spatial resolution. This gives the possibility to obtain velocity dispersion 
of the galaxies which, together with luminosity, can be used to calculate the 
M/L ratio of galaxies that is essential information for tracing the young population. Furthermore, the 
correlation between velocity dispersion and K-band NaI index can be used as a young population indicator for early-type 
galaxies as was suggested by \citet{silv08}. To characterize stellar population gradients from the central region, 
where AGN and star formation may co-exist, to the outer regions of galaxies higher spectral and spatial resolution are needed. 
Furthermore, to better interpret the composite NIR spectra of galaxies, more extended stellar population synthesis models 
should be developed. To assert the unified model of radio-loud AGNs, where 
RGs are the parent population of BL Lac objects, larger samples of NIR spectra of BL Lac objects are 
required together with high resolution optical spectroscopy.


\begin{acknowledgements}
This work was supported by the Italian Ministry for University and Research (MIUR) 
under COFIN 2002/27145, ASI-IR 115 and ASI-IR 35, ASI-IR 73 
and by the Academy of Finland (projects 8121122 and 8107775). 
This research has made use of the NASA/IPAC Extragalactic Database {\em(NED)} which is operated by 
the Jet Propulsion Laboratory, California Institute of Technology, 
under contract with the National Aeronautics and Space Administration. 
We thank the referee, Alexandre Vazdekis, for his constructive criticism, 
which significantly improved the paper, and Valentin Ivanov for allowing us 
to use the data on inactive elliptical galaxies prior to publication.

\end{acknowledgements}

\bibliographystyle{aa} 
\bibliography{totahy.bib}

\end{document}